\documentclass[lettersize,journal]{IEEEtran}
\usepackage{amsmath,amsfonts}
\usepackage{algorithmic}
\usepackage{algorithm}
\usepackage{array}
\usepackage[caption=false,font=normalsize,labelfont=sf,textfont=sf]{subfig}
\usepackage{textcomp}
\usepackage{stfloats}
\usepackage{url}
\usepackage{verbatim}
\usepackage{graphicx}
\usepackage{cite}
\usepackage{setspace}
\usepackage[table]{xcolor}
\usepackage{multirow}
\usepackage{booktabs}
\usepackage{rotating}
\usepackage[hidelinks]{hyperref}
\usepackage[T1]{fontenc}
\usepackage{tikz}
\usetikzlibrary{positioning, arrows.meta}
\begin{document}

\title{Energy-Efficient GPU DVFS for Fine-Tuning of SLMs on Resource-constrained Embedded Devices}


\author{Jurn-Gyu Park, 
Sanzhar Zholdybayev, 
Aidar Amangeldi, and 
Ademi Zhanuzakova

\thanks{This manuscript is currently under review for publication; (Corresponding author: jurn.park@nu.edu.kz.)}
\thanks{The authors gratefully acknowledge Prof. Atakan Varol for providing access to the computational resources of the Institute of Smart Systems and Artificial Intelligence (ISSAI), and Yerbol Absalyamov and Vladimir Albrekht for their assistance.}
\thanks{The authors are with the School of Engineering and Digital Sciences,
Nazarbayev University, 010000 Astana, Kazakhstan
(e-mails: jurn.park,sanzhar.zholdybayev,aidar.amangeldi,ademi.zhanuzakova@nu.edu.kz).}
}

\markboth{}%
{Shell \MakeLowercase{\textit{et al.}}: A Sample Article Using IEEEtran.cls for IEEE Journals}


\maketitle
\begin{spacing}{0.9}
\begin{abstract}
Dynamic Voltage Frequency Scaling (DVFS) on resource-constrained embedded GPU platforms is essential for energy-efficient small language model (SLM) fine-tuning, as privacy- and personalization-driven adaptation increasingly requires local execution and involves repeated forward-backward optimization over many mini-batches, making it substantially more time- and energy-intensive than single-pass inference. 
%
To this end, 1) we first characterize the fine-tuning behavior of representative encoder-only SLMs of BERT variants, and autoregressive decoder-only SLMs of Pythia variants on GLUE benchmarks. In addition to the characterizations, 2) we propose a simple yet effective ML-based model selection that selects energy-optimal GPU DVFS settings on resource-constrained embedded platforms.
Our results on NVIDIA Jetson AGX Orin demonstrate average 13.11\% energy savings (up to 26.73\%) over MAXN Mode 0, which has no explicit power cap. 
\end{abstract}





\section{Introduction}


\IEEEPARstart{T}{ransformer} models on integrated GPU-based embedded and mobile platforms are increasingly adopted for on-device SLM fine-tuning in healthcare, finance, and autonomous systems, where privacy, latency, and offline operation requirements forbid cloud offload~\cite{edge_llm_review}~\cite{deploying_ai_edge}.
These battery-based and resource-constrained devices require crucial yet challenging dynamic power management (DPM) techniques for power and energy savings without accuracy degradation, using power modes and/or dynamic voltage frequency scaling (DVFS) on heterogeneous processors.

A line of studies focuses on DVFS for large language model (LLM) inference workloads~\cite{throttLLeM2024,Camel2025,Maliakel2025} and datacenter training~\cite{Zeus2023,Perseus2024,KernelDVFS2026}; moreover, the off-the-shelf Jetson AGX Orin series support four different power modes: 15W, 30W (Default/Mode 2), 50W, and the unconstrained MAXN performance mode (Mode 0). 
To the best of our knowledge, our work firstly aims energy-efficient GPU DVFS for small language models (SLMs) fine-tuning using encoder-based and decoder-based SLM models on Jetson AGX Orin (64GB) unconstrained MAXN Mode, with the help of comprehensive application-specific workloads characterization.
The contributions of the paper are as follows:

\begin{itemize}
  \item We characterize how GPU frequency scaling affects fine-tuning latency, power consumption, and total energy on Transformer-based SLM workloads. 
  

  \item We propose lightweight GPU frequency governors for energy-efficient SLM fine-tuning based on interpretable ML-based DT model policy.  
  
  
  \item We demonstrate up to $26.73\%$ energy savings on NVIDIA Jetson AGX Orin accross benchmark fine-tuning workloads. 

  
\end{itemize}

\section{Motivation and Related Work}



\bgroup
\begin{figure}[t]
\centering
\includegraphics[width=0.95\columnwidth]{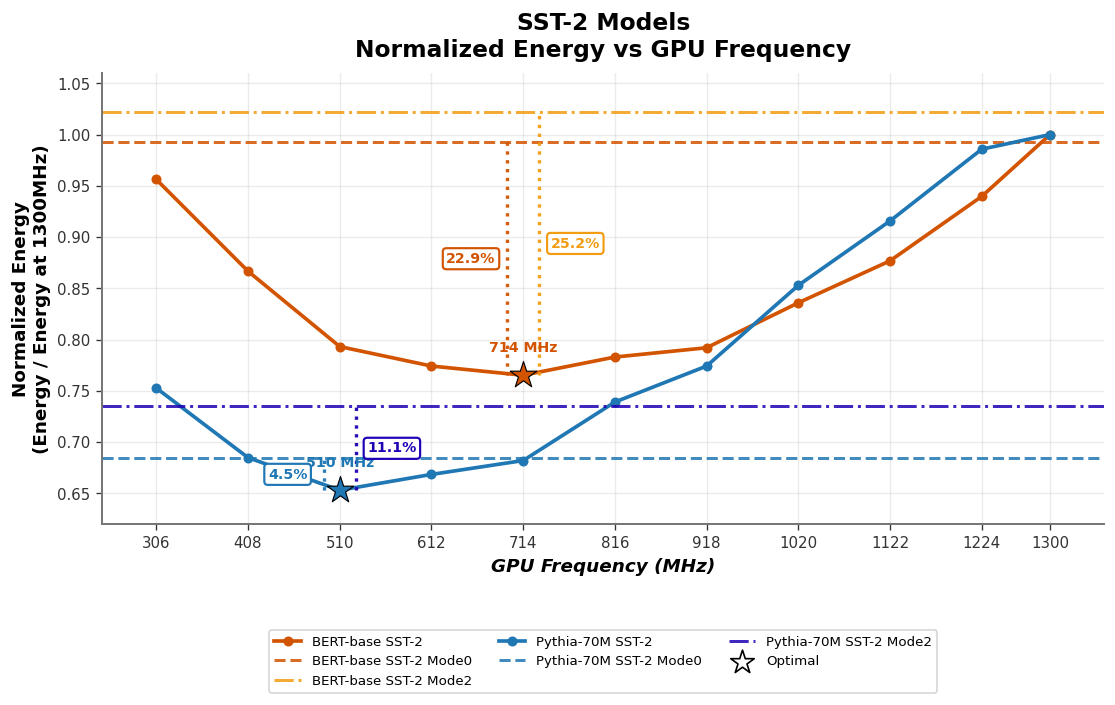}
\caption{Motivating example: Normalized fine-tuning energy versus GPU frequency for BERT-base and Pythia-70M on SST-2 using Jetson AGX Orin. 
}
\label{motivating-example}
\end{figure}
\egroup

\subsection{Motivation}


NVIDIA Jetson platforms~\cite{NVIDIA_Jetson_Benchmarks} provide predefined power modes~\cite{jetson_power_performance_documentation} 
for power-aware embedded deployment. However, these coarse-grained profiles do not guarantee energy-efficient fine-tuning, since total energy depends on both power draw and execution time. This issue is critical for on-device SLM adaptation, where repeated forward–backward optimization must run under limited compute, memory, and battery resources. 
Motivated by this gap, we characterize effects of GPU-frequency behavior during encoder- and decoder-only SLM fine-tuning, and propose a lightweight sweet-spot policy for selecting energy-efficient GPU DVFS settings on embedded platforms.
%

We sweep the GPU frequency over 11 discrete levels from 306 MHz to 1300 MHz, with CPU frequency managed by the default Jetson governor. On the GLUE benchmark, all tested configurations retain identical validation accuracy, but show distinct latency–power trade-offs. As illustrated in Figure~\ref{motivating-example}, the minimum-energy point is workload-dependent and follows a U-shaped pattern, achieving lower energy than both the default 30 W-capped Mode 2 and the unconstrained MAXN Mode 0.
\IEEEpubidadjcol
This motivating example demonstrates that static Jetson power modes can operate away from the energy-minimizing point, resulting in up to $25$\% avoidable energy consumption that can be eliminated through workload-aware GPU frequency selection.

\begin{figure*}[t]
\centering
\includegraphics[width=0.99\textwidth]{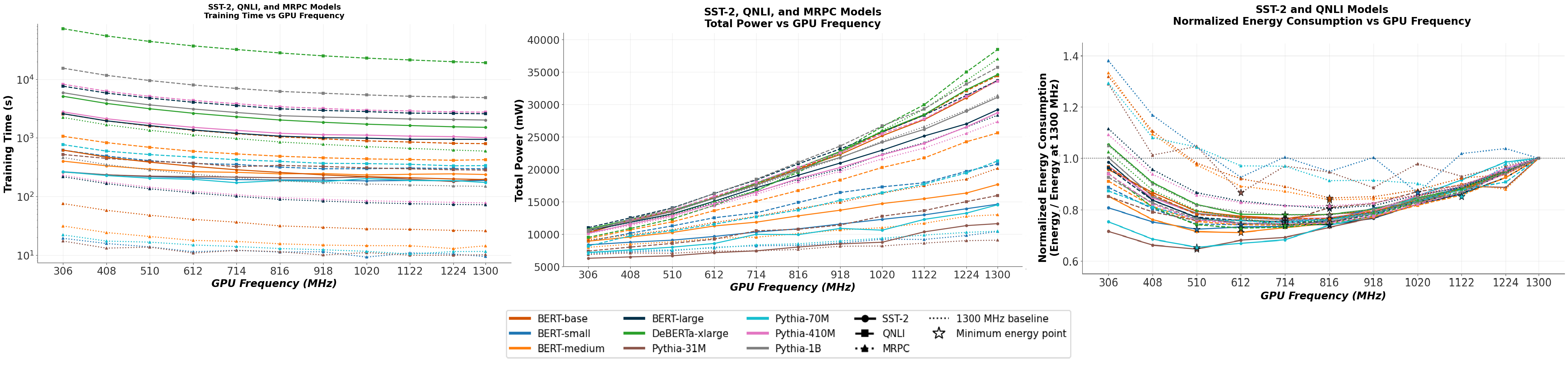}
\caption{
Fine-tuning time, total power, and energy versus GPU frequency for BERT and Pythia variants on SST-2, QNLI, and MRPC (GLUE). [Left] Frequency vs. Fine-tuing time correlation. [Middle] Frequency vs. Total power correlation. [Right] Frequency vs. Energy correlation. 
}

\label{characterization}
\end{figure*}

\subsection{Related Work}
Some recent works on GPU DVFS for LLM inference on datacenter platforms~\cite{Maliakel2025}~\cite{GreenLLM2025}, embedded inference~\cite{Camel2025}~\cite{DVFO2024},
training-focused datacenter studies using power capping and pipeline-aware scheduling~\cite{Zeus2023}~\cite{Perseus2024}~\cite{EnvPipe2023}~\cite{KernelDVFS2026},
and integrated CPU-GPU DVFS studies on embedded systems for CNN workloads~\cite{Karzhaubayeva2023},
transformer fine-tuning workloads characterization for GPU DVFS studies on resource-constrained embedded platforms are very rare.
A recent study of federated fine-tuning on Jetson AGX Orin~\cite{Woisetschlager2024} operated exclusively under the default power mode without varying GPU frequency, and characterizations across Jetson generations~\cite{Prashanthi2025,Prashanthi2023} focus on inference or coarse-grained NVP-mode comparisons rather than per-frequency energy curves for fine-tuning.




\section{Methodology}
As shown in Figure~\ref{methodology-overview}, our methodology is composed of five phases: 1) data collection, 2) SLM workloads characterization, 3) ML model selection, 4) design for governor policies, and 5) policy evaluation.

\bgroup
\begin{figure}[h]
\centering
\resizebox{\columnwidth}{!}{
\begin{tikzpicture}[
    box/.style={
        draw,
        rectangle,
        rounded corners=2pt,
        minimum height=0.55cm,
        minimum width=1.35cm,
        align=center,
        font=\tiny
    },
    arrow/.style={
        -{Latex[length=1mm]},
        line width=0.2pt
    },
    node distance=0.18cm
]

\node[box] (data) {Data Collection}; 
\node[box, right=of data] (slm) {SLM Workloads\\Characterization};
\node[box, right=of slm] (ml) {ML Model\\Selection};
\node[box, right=of ml] (gov) {Governor\\Policies Design};
\node[box, right=of gov] (eval) {Policy \\Evaluation};

\draw[arrow] (data) -- (slm);
\draw[arrow] (slm) -- (ml);
\draw[arrow] (ml) -- (gov);
\draw[arrow] (gov) -- (eval);

\end{tikzpicture}
}

\caption{Methodology Overview.
}
\label{methodology-overview}
\end{figure}
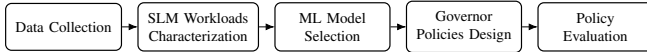
\egroup

\subsection{Experimental Setup and Data Collection}


\subsubsection{Experimental Setup}
\textbf{Platform}: NVIDIA Jetson AGX Orin Developer Kit ~\cite{jetson_agx_documentation} with $64\,$GB LPDDR5 ($204.8\,$GB/s bandwidth), Ampere-architecture GPU with $2048$ CUDA cores (CUDA $12.6$) and $64$ Tensor Cores, and $12$-core ARM Cortex-A78AE CPU.
The platform supports $11$ discrete GPU frequencies from $306$ to $1300\,$MHz under user-space governor control.
\textbf{SLMs}: Encoder-only BERT
~\cite{bert} variants: BERT-small($29$M, $4$ layers), BERT-medium($41$M, $8$ layers), BERT-base ($110$M, $12$ layers), BERT-large($340$M, $24$ layers), and DeBERTa-v2-xlarge ($900$M, $24$ layers). Decoder-only Pythia~\cite{pythia} variants: Pythia-31M ($31$M, $6$ layers), Pythia-70M ($70$M, $6$ layers), Pythia-410M ($410 $M, $24$ layers), Pythia-1B ($1$B, $16$ layers). 
HuggingFace \texttt{transformers} (PyTorch $2.5.0$) are adopted. 
\textbf{Benchmark}: Three tasks from GLUE benchmark~\cite{wang2019glue} of SST-2 (sentiment, $67$K examples), QNLI (question NLI, $104$K examples), and MRPC (paraphrase, $5.8$K examples) are used for characterization and model selection phase, while additional unseen tasks (QQP, RTE, CoLA, STS-B, MNLI, WNLI) are utilized for test phase.

\subsubsection{Data Collection and Fine-Tuning}
\textbf{Data Collection}: 
To collect data, the \emph{tegrastats} tool~\cite{NVIDIA_Jetson_Benchmarks} at $10\,$ms sampling intervals and a custom parser utility in Python were used. We recorded data on training time, CPU and GPU power consumption (total power), as well as CPU, external memory controller (EMC), and GPU frequency and utilizations. These experiments were conducted on NVIDIA Jetson AGX Orin using Transformer based BERT and Pythia variants using GLUE benchmark. Available GPU frequencies (details in Table~\ref{available-freqs}) are swept, using the default CPU governor \emph{schedutil}~\cite{jetson_power_performance_documentation} with 29 available CPU frequencies in range of $120$-$2200$ MHz.
%
%
\textbf{Fine-Tuning}: Fine-tuning was profiled for each configuration with two methods: fixed duration of 5 minutes and fixed number of steps. Since each model has different fine-tuning speed, the number of steps was selected individually for each model and task based on the total number of training steps and the per-step execution time. For example, on the QQP task, 100 steps takes approximately $350$s for Pythia-1B, but only about $10$s for Pythia-31M. For the final evaluation, 5-minute method was used because it provided more stable and fair numerical results across all experiments. However, if running time for full-fine tuning was less than five minutes, the results of full-fine tuning were used instead. The hyperparameters the same across the runs: $1$ epoch, AdamW optimizer, learning rate $2{\times}10^{-5}$, FP16 mixed precision, and fixed seed. The exception was DeBERTa-xlarge, which required a lower learning rate of $1{\times}10^{-5}$ and full FP32 precision to prevent numerical instabilities.

\begin{table}[t]
\caption{Available GPU frequencies (MHz). \label{available-freqs}
}
\centering
\scriptsize
\begin{tabular}{p{\dimexpr.077\linewidth-2\tabcolsep}p{\dimexpr.077\linewidth-2\tabcolsep}p{\dimexpr.077\linewidth-2\tabcolsep}p{\dimexpr.077\linewidth-2\tabcolsep}p{\dimexpr.077\linewidth-2\tabcolsep}p{\dimexpr.077\linewidth-2\tabcolsep}p{\dimexpr.077\linewidth-2\tabcolsep}p{\dimexpr.077\linewidth-2\tabcolsep}p{\dimexpr.077\linewidth-2\tabcolsep}p{\dimexpr.077\linewidth-2\tabcolsep}p{\dimexpr.077\linewidth-2\tabcolsep}p{\dimexpr.077\linewidth-2\tabcolsep}}
\hline
Idx. & 0 & 1 & 2 & 3 & 4 & 5 & 6 & 7 & 8 & 9 & 10 \\
\hline
GPU & 306 & 408 & 510 & 612 & 714 & 816 & 918 & 1020 & 1122 & 1224 & 1300 \\
\hline
\end{tabular}
\end{table}


\subsection{Characterization for Latency–Power–Energy Trade-offs}

%
%
Figure~\ref{characterization} represents the correlation between GPU frequency and fine-tuning time, total power, plus total energy on BERT variants (BERT-small/medium/base/large and DeBERTa-xlarge) and Pythia variants (Pythia-31M, Pythia-70M, Pythia-410M, and Pythia-1B) on SST-2, QNLI, and MRPC of GLUE.


\textbf{Frequency–Latency Correlation}: Figure~\ref{characterization} [Left] represents the relationship between GPU frequencies and fine-tuning time. The y-axis is the execution time in seconds, which almost exponentially decreases as the GPU frequency increases. 

\textbf{Frequency–Total Power Correlation}: Figure~\ref{characterization} [Middle] represents the relationship between GPU frequencies and total power consumption (CPU + GPU). The relation of frequency to power is proportional according to $P \propto C \cdot V^{2} \cdot f$. Power increases as GPU frequencies increase. The highest GPU frequencies often consume more power than the Mode0 baseline.

\textbf{Frequency–Energy Correlation}: Figure~\ref{characterization} [Right] represents the relationship between GPU frequencies and total energy consumption. The energy is normalized to the maximum available frequency ($1300$MHz) using $E_{\mathrm{norm}}(f) = \frac{E(f)}{E(1300\,\mathrm{MHz})}$. The lines exhibit a well-defined U-shape: energy increases at low frequencies due to extended training time, and at high frequencies due to elevated power consumption. The optimal points are mostly in the range between $510$-$816$MHz. 
However, the challenge question is \textit{how to select the most energy-efficient frequencies for various SLM workloads?} To do this, we propose ML model based policies. 
\subsection{ML Model Selection}

\textbf{Dataset}:
It is collected from 27 model-task pair configurations and is composed of five selected independent features: model type, GPU utilization, number of layers, hidden size, and average sequence length; and the target variable is the optimal GPU frequency level, encoded as labels from 1 to 6 (i.e., 510, 612, 714, 816, 1020, and 1122 MHz). The dataset is imbalanced, and the largest class/label corresponds to the optimal GPU frequency of 714 MHz.

\begin{figure}[H]
\centering
\includegraphics[width=\linewidth]{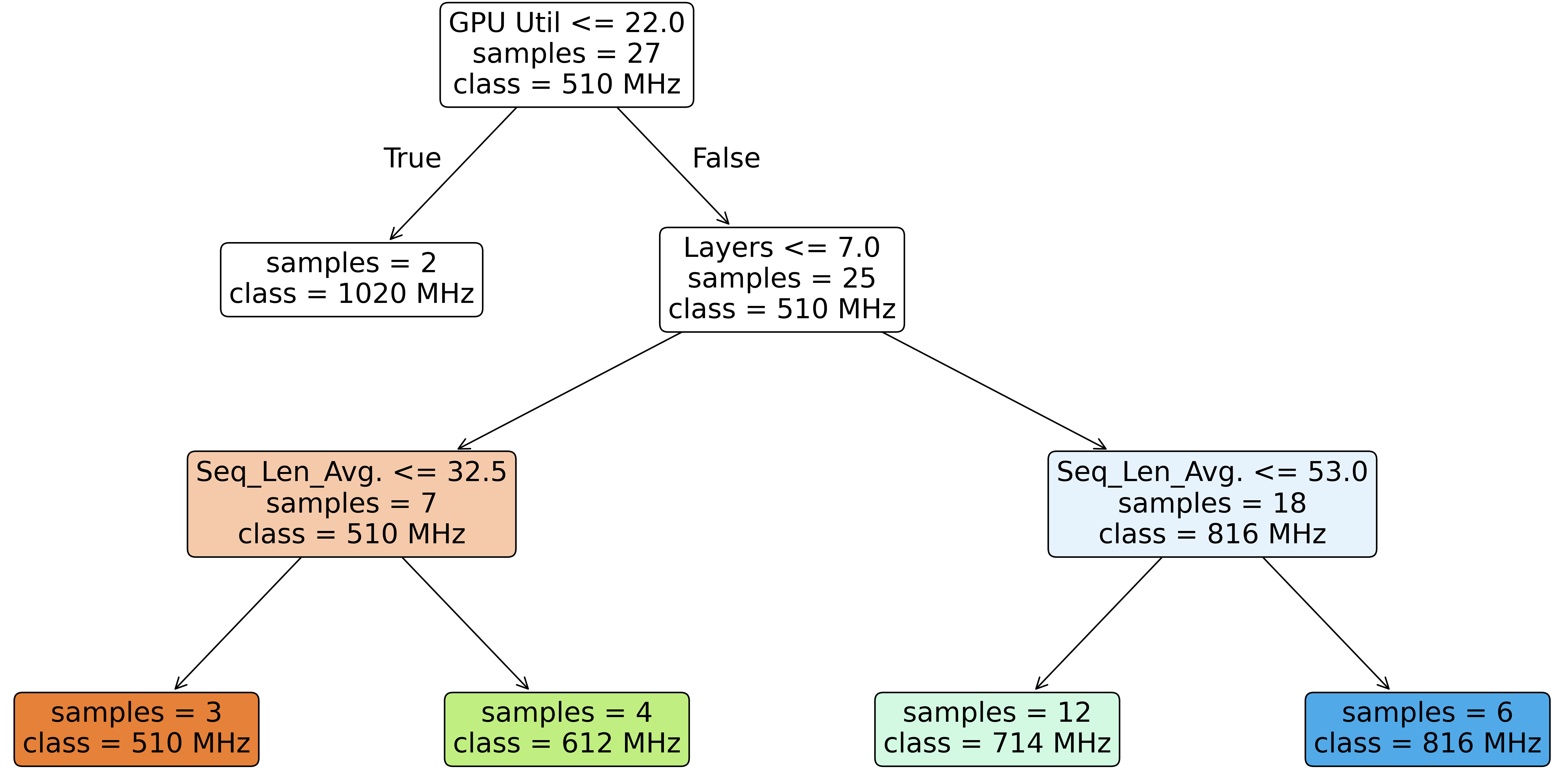}
\caption{Interpretable Tree structure of the selected model . 
} 
\label{fig:dttree}
\end{figure}

\textbf{Model Selection}:
The most energy-efficient GPU-frequency prediction task is formulated as a multi-class classification problem. The Decision Tree (DT) classifier is selected because of its inherent interpretability and ability to map these features to GPU frequency labels. 
Considering the limited number of samples available, the hyperparameters were chosen based on grid search with the exhausitive Leave-One-Out cross-validation (LOOCV) to maximize utility of the training set. The DT model achieved a validation accuracy of 0.62 with the interpretable tree structure in Figure~\ref{fig:dttree}. 

Although the performance of the model is moderate, it gives a transparent and deterministic decision rule to derive the GPU-frequency policy, remaining as future work for more accurate DT models. 



\subsection{ML model based Governor Policy}


\begin{algorithm}
\small
\caption{Interpretable DT Model based GPU Governor 
}\label{dt-algorithm}
\begin{algorithmic}[1]
\REQUIRE  Workload (model $M$, dataset $D$)
\ENSURE Energy-efficient GPU frequency $f^{*}$
\IF{$Layers \leq 7$}
    \IF{$SeqLenAvg\leq 32.5$}
        \STATE $f^{*} \leftarrow 510\,$MHz
    \ELSE
        \STATE $f^{*} \leftarrow 612\,$MHz
    \ENDIF
\ELSE
    \IF{$SeqLenAvg \leq 53.0$}
        \STATE $f^{*} \leftarrow 714\,$MHz
    \ELSE
        \STATE $f^{*} \leftarrow 816\,$MHz
    \ENDIF
\ENDIF
\IF {$U_{\mathrm{GPU}} \leq 22.0$}
    \STATE $f^* \leftarrow 1020\,\mathrm{MHz}$
\ENDIF
\end{algorithmic}
\end{algorithm}


As shown in Algorithm~\ref{dt-algorithm}, an interpretable DT model based policy is designed with the structural features: transformer depth (\textit{Layers}) and sequence length (\textit{SeqLenAvg}). By utilizing the above features, the governor generates clear deterministic policies.
Specifically, for models with smaller depths ($\textit{Layers} \le 7$), the average sequence length determines the choice of frequency: shorter sequences ($SeqLenAvg \leq 32.5$) map to $510\,$MHz, while longer sequences assigned to $612\,$MHz. For models with greater depths ($\textit{Layers} > 7$), the sequence-length decision threshold becomes $53.0$, based on which they are classified to $714\,$MHz or $816\,$MHz, respectively.

\subsection{Evaluation}


Finally, we evaluated improvements in energy consumption compared to Mode0 (unconstrained dynamic frequency scaling up to $1300\,$MHz). The algorithm is tested on unseen workloads. 


\section{Experimental Results and Analysis}




Table~\ref{dt-mode0-results} presents results across six unseen test workloads (QQP, RTE, CoLA, STS-B, MNLI, WNLI) using BERT and Pythia variants, compared to Mode0. 
In the experiments, we compare two governor policies: 1) Majority rule based fixed frequency governor (FFG), and 2) the interpretable ML (DT) model based governor (MLG), described in Algorithm~\ref{dt-algorithm}.

\subsubsection{Results of the Fixed frequency governor (FFG)} 

Although larger models in the fixed-frequency setting have stable improvements above $17\%$ across frequencies, small-scale models ($\leq70$M parameters) show uncertainty. For example, BERT-medium shows $18.38$\% on QQP and $-2.31$\% on CoLA. 
This happens because of the fluctuating pattern of the small models. An unclear U-shaped pattern in the Energy and GPU frequency graph may shift the optimal point far from $714$MHz. For example, BERT-small and Pythia-70M on MRPC have optimal points at $1020$MHz and $1122$MHz, respectively. Another reason is spikes in the curve. For example, Pythia-31M on MRPC has an optimal point on $612$MHz; however, at $714$MHz there is a sudden and dramatic increase in energy, making it worse than Mode0 by $5$\%. We speculate that the number of parameters is important for producing a clear U-shaped pattern and stable results. Overall, the average result is better by $12.21$\% compared to Mode0


\subsubsection{Results of the interpretable ML model based governor (MLG)} 

Similar to the fixed-frequency policy, larger models show higher and more stable improvements, with average savings of larger models ($>70$M parameters) greater than $22$\%. 
 In contrast, smaller models still have uncertain results. For example, Pythia-70M shows a $9.38$\% improvement on QQP and $-7.23$\% on WNLI. These results strengthen the speculation that the number of parameters affects result stability. Since larger models exhibit a clearer U-shaped curve, their results show better energy savings than Mode0.

\subsubsection{Comparison between FFG and MLG policies} 

Compared to the FFG policy, the MLG policy improves the final average by $0.9\%$, which is mainly affected by the $3.54\%$ improvement on the RTE task. This suggests the MLG policy may better capture the behavior of longer-sequence workloads. The QQP and WNLI tasks show small positive differences of $0.22\%$ and $0.10\%$, respectively, but the CoLA task shows a degradation of $0.54\%$. This small degradation is caused mainly by the GPU utilization branch routing some CoLA cases to a higher frequency than their optimal value. The MLG policy improves the average result for each model compared to the FFG policy. The largest improvements are observed for the two smallest models: Pythia-31M and BERT-small with improvements of $2.53$\% and $2.29$\%, respectively. Smaller models have dramatic changes in some tasks, including changes from negative to positive values and vice versa. For example, Pythia-31M increased from $-4.15$\% to $8.71$\% on RTE, while simultaneously decreasing from $9.82$\% to $-0.35$\% on CoLA. The larger models show only slight increases or no change, indicating a more stable U-shaped energy–frequency curve. This further supports the speculation that parameter count directly scales result stability.

\begin{table*}[!t]
\caption{\scriptsize  Comparison of FFG and MLG policies, against Mode0.\label{dt-mode0-results}
}
\centering
\scriptsize
\setlength{\tabcolsep}{3pt}
\renewcommand{\arraystretch}{0.95}

\begin{tabular}{c|cccccc|cccccc|cc}
\hline
\multirow{2}{*}{Model} &
\multicolumn{6}{c|}{$\Delta E_{\mathrm{FFG}}$ (\%)} & 
\multicolumn{6}{c|}{$\Delta E_{\mathrm{MLG}}$ (\%)} & 
\multicolumn{2}{c}{Model average}\\

& QQP & RTE & CoLA & STS-B & MNLI & WNLI
& QQP & RTE & CoLA & STS-B & MNLI & WNLI
& {$\Delta E_{\mathrm{FFG}}$} & {$\Delta E_{\mathrm{MLG}}$}\\
\hline

BERT-small 
& 2.49  & -7.27 & -9.44 & -2.94 & -3.12 & -1.94
& 3.00  & -1.49 & -8.63 & -0.55 & -0.31 & -0.53 
& -3.7 & -1.42\\
BERT-medium
& 18.38 & 0.48  & -2.31 & 9.49 & 4.56 & 1.45 
& 18.38 & 6.90  & -2.31 & 9.49 & 4.56 & 1.45
& 5.34 & 6.41\\
BERT-base 
& 24.37 & 9.38  & 16.65 & 22.22 & 24.50 & 26.73
& 24.37 & 12.37 & 16.65 & 22.32 & 24.05 & 25.80 
& 20.64 & 20.93\\
BERT-large
& 24.91 & 17.58 & 20.95 & 23.69 &  24.51 & 23.44 
& 24.91 & 18.94 & 20.95 & 23.69 & 24.51 & 23.44 
& 22.51 & 22.74\\
DeBERTa-xlarge 
& 22.83 & 22.86 & 22.82 & 22.39 & 25.68 & 23.27 
& 22.83 & 23.10 & 22.82 & 22.39 & 25.68 & 23.27 
& 23.31 & 23.35\\
\hline
Pythia-31M
& -9.72 & -4.15 & 9.82  & -9.96 & -5.23 & -3.05 
& -7.66 & 8.71  & -0.35 & -6.79 & -0.41 & -0.64 
& -3.72 & -1.19\\
Pythia-70M
& 6.79  & -1.34 & -1.57 & 6.41 & -3.97 & -5.21 
& 6.22  & -0.52 & 2.94 & 9.38 & -0.98 & -7.23 
& 0.19 & 1.64\\
Pythia-410M
& 24.43 & 17.88 & 19.89 & 23.47 & 24.62 & 22.67
& 24.43 & 18.86 & 19.89 & 23.47 & 24.62 & 22.67
& 22.16 & 22.32\\
Pythia-1B
& 24.98 & 20.53 & 21.54 & 24.31 & 25.38 & 22.27
& 24.98 & 20.98 & 21.54 & 24.31 & 25.38 & 22.27 
& 23.17 & 23.24\\
\hline
Task average 
& 15.50 & 8.44 & 10.93 & 13.23 & 12.99 & 12.18
& 15.72 & 11.98 & 10.39 & 14.19 & 14.12 & 12.28 \\
\hline
Average &\multicolumn{6}{c|}{12.21}&\multicolumn{6}{c|}{13.11}\\
\hline
\end{tabular}
\\[2pt]
{\scriptsize 
$\Delta E_{FFG} (\%) =
(E_{\mathrm{Mode0}} - E_{FFG})/E_{\mathrm{Mode0}} \times 100 (\%)$ and
$\Delta E_{\mathrm{MLG}} (\%) =
(E_{\mathrm{Mode0}} - E_{\mathrm{MLG}})/E_{\mathrm{Mode0}} \times 100 (\%)$. \\
Positive values indicate energy savings over Mode0; negative values indicate higher energy consumption.
}
\end{table*}




\section{Conclusion and Future Work}

The paper presents a comprehensive characterization of transformer fine-tuning workloads using BERT and Pythia SLM applications on Jetson AGX Orin, and
proposes an ML-enhanced frequency selection policy for energy-efficient fine-tuning, compared with the off-the-shelf power modes.
Our FFG and MLG achieve average $13.11\%$ (up to $26.73\%$) energy improvements on unseen validation workloads versus the off-the-shelf modes. 
%

Although the effects of FFG or MLG policies are clear, our work has limitations of needs of more diverse SLM and benchmark workloads, having higher accuracy with the help of more sophisticated ML models and MLG policies to achieve more energy savings compared to the FFG policy, which is remaining as a future work. 


\bibliographystyle{IEEEtran}
\bibliography{main}







\end{spacing}
\end{document}